\documentclass[aps,prl,twocolumn,showpacs,preprintnumbers]{revtex4}
\usepackage{amsmath}
\usepackage{dcolumn}
\usepackage{bm}
\usepackage{subfigure}
\usepackage{amsfonts}
\usepackage{amssymb}
\usepackage{makeidx}
\usepackage{epsfig}
\usepackage{graphicx}

\begin{document}

\title{\textbf{Lagrangian dynamics of thermal tracer particles\\
in Navier-Stokes fluids}}
\author{Massimo Tessarotto\thanks{%
Electronic-mail: M.Tessarotto@cmfd.univ.trieste.it}$^{a,b}$, Claudio
Cremaschini$^{c}$ and Marco Tessarotto$^{a,d,e}$}
\affiliation{$^{a}$Dept.of Mathematics and Informatics,University of Trieste, Italy \\
$^{b}$Consortium of Magnetofluid-dynamics, University of Trieste, Italy\\
$^{c}$International School for Advanced Studies (SISSA) and INFN, Trieste,
Italy\\
$^{d}$Dept.of Electronics, Electrotechnics and Informatics, University of
Trieste, Italy \\
$^{e}$Civil Protection Agency, Regione Friuli Venezia-Giulia, Palmanova
(GO), Italy}
\date{\today }

\begin{abstract}
A key issue in fluid dynamics is the definition of the phase-space
Lagrangian dynamics characterizing prescribed ideal fluids (i.e., continua),
which is related to the dynamics of so-called \textit{ideal tracer particles}
moving in the same fluids. These are by definition particles of
infinitesimal size which do not produce significant perturbations of the
fluid fields and do not interact among themselves. In this work we point out
that the phase-space Lagrangian description of incompressible Navier-Stokes
fluids can be achieved by means of a particular subset of ideal tracer
particles, denoted as thermal particles. For these particles the magnitude
of their relative velocities - with respect to the local fluid velocity - is
solely determined by the kinetic pressure, in turn, uniquely related to the
fluid pressure. The dynamics of thermal tracer particles is shown to
generate the time-evolution of the fluid fields by means of a suitable
statistical model. The result is reached introducing a 1-D statistical
description of the fluid exclusively based on the ensemble of thermal tracer
particles. In particular, it is proven that the statistic of thermal
particles can be uniquely defined requiring that for these particles the
directions of their initial relative velocities are defined by a suitable
family of random coplanar unit vectors.
\end{abstract}

\maketitle


\section{1 - Introduction}

The phase-space Lagrangian description of incompressible Navier-Stokes (NS)
fluids is an essential ingredient of turbulence theory \cite%
{Pope2000,Tessarotto2009A}. Turbulence, in fact, can be viewed as the
manifestation of stochastic behavior of the underlying phase-space dynamical
system which advances in time the state of the fluid (the so-called \textit{%
NS dynamical system} \cite{Ellero2005}). It is therefore important to
provide an insight into the phase-space dynamics which characterizes these
systems \cite{Cremaschini2008,Tessarotto20083,Tessarotto2009}.

In the framework of the theory of continua, for a given fluid system, there
is a unique minimal class of such dynamical systems which provide - via the
introduction of an appropriate phase-space probability density function
(PDF) - the time evolution of the complete set of fluid fields describing
the same fluid. In turbulence theory this phase space is usually identified
with the set $\Gamma =\Omega \times U,$ with $dim(\Gamma )=6$, denoted as
\emph{restricted phase-space} (with $U\equiv
\mathbb{R}
^{3}$ denoting the velocity space and $\overline{\Gamma }=\overline{\Omega }%
\times U$ the closure of $\Gamma $ ). Therefore it is natural to seek
possible phase-space representations of this type for fluid systems.

Although phase-space descriptions of incompressible NS fluids have been
around for a long time, starting from the historical work of Hopf (Hopf,
1952 \cite{Hopf1952}), Edwards (Edwards, 1964 \cite{Edwards1964}) and Rosen
(Rosen, 1971 \cite{Rosen1971}), a\ fundamental issue concerns the
construction of phase-space approaches exclusively based on \emph{classical
statistical mechanics} (CSM). This concerns, in particular, the search of
so-called \textit{inverse kinetic theories} (IKT) able to yield the complete
set of fluid fields describing the fluid. In these approaches, denoted as
\emph{complete IKT's} \cite{Tessarotto20082}\emph{,} the fluid is described
in terms of a statistical ensemble of point-particles, called \emph{ideal}
\emph{tracer particles}, whose phase-space trajectories are defined by the
general solution\ of the initial-value problem of the form%
\begin{equation}
\left\{
\begin{array}{c}
\frac{d}{dt}\mathbf{r}(t)=\mathbf{v}(t), \\
\frac{d}{dt}\mathbf{v}(t)=\mathbf{F}(\mathbf{r}(t),t;f), \\
\mathbf{r}(t_{o})=\mathbf{r}_{o}, \\
\mathbf{v}(t_{o})=\mathbf{v}_{o},%
\end{array}%
\right.  \label{EQ.0}
\end{equation}%
where $\mathbf{F}$\textit{\ }is a suitable vector field (\emph{mean-field
force}) and the state-vector $\mathbf{x}=(\mathbf{r,v})$ spans by assumption
the phase space $\Gamma \equiv \Omega \times U.$ For them, as pointed out
elsewhere \cite{Tessarotto20083,Tessarotto2009} and unlike traditional
approaches \cite{Tchen1947,Corrsin1956,Buevich1966,Riley 1971,Maxey}, the
dynamics (defined in terms of\ $\mathbf{F}$) can be in principle uniquely
established based exclusively on first principles. As a result, by
construction, in a complete IKT, ideal tracer particles:

\begin{itemize}
\item uniquely determine, by means of the 1-point statistics, the time
evolution of the fluid fields;

\item interact with the fluid via a suitable mean-field force $\mathbf{F}$;

\item do not perturb the fluid fields;

\item are collisionless (i.e., they do not interact among themselves);

\item are subject to a mean-field force defined in such a way that the
dynamics of the ensemble of tracer particles reproduces exactly, via a
suitable statistics, the dynamics of the fluid fields. In particular, the
complete set of fluid fields is determined in terms of the 1-point PDF.
\end{itemize}

An IKT of this type, in which tracer particles can have \emph{arbitrary
initial velocities} (i.e., which belong to the whole velocity space $U\equiv
\mathbb{R}
^{3}$), has been recently established for incompressible thermofluids (see
in particular, Tessarotto et al., \cite{Tessarotto2009}). In particular, in
the case of an incompressible and isothermal NS fluid it is immediate to
show that the dynamics of tracer particles generally differs from that of
the corresponding fluid elements \cite%
{Tessarotto20083,Tessarotto2009,Tessarotto20081}. The difference is
intrinsic, i.e., it is produced by the mean-field force acting on the tracer
particles (which differs from the local force density acting on the fluid
elements), so that generally the fluid velocity is \emph{not} a solution of
the NS dynamical system.

A peculiar characteristics of a complete IKT is, however, that the
mean-field force $\mathbf{F}$ contains an arbitrary scalar parameter ($a$)
\cite{Tessarotto2007a}$.$ This feature, which - nevertheless - is not
reflected by the fluid fields (which remain by construction independent of
the same parameter), can also be interpreted as the non-uniqueness in the
definition of the NS dynamical system.

An interesting question is, however, whether a proper subset of solutions
(of the NS dynamical system) exists which can be used to characterize
uniquely incompressible NS fluids and resolve also uniquely the previous
indeterminacy issue. This refers, in particular, to the possibility of
identifying a proper subset of the family of tracer particles, characterized
by a phase-space of lower dimension (i.e., smaller than $6$) and whose
states uniquely determine \emph{exactly }- via a suitable statistics\ - both
the complete set of fluid fields and their time evolution.

\

\subsection{Goals of the investigation}

Purpose of this paper is to point out the discovery of a family of tracer
particles denoted as \emph{thermal tracer particles, }which have all such
features. These tracer particles are characterized by relative velocities
belonging to a subset of the velocity space $U$ of dimension 1. As a main
consequence, it is shown (see THM.1) that the dynamics of these particles
uniquely determines - via a suitable statistics - the time-evolution of the
NS fluid. The ensemble of these particles is defined by the requirement that
for each tracer particle the magnitude of its relative velocity (with
respect to the local fluid velocity) is proportional to the fluid pressure.
More precisely, denoting by $\mathbf{x}(t)=\left\{ \mathbf{r}(t),\mathbf{v}%
(t)\right\} $ the phase-space trajectory of a generic thermal tracer
particle, with $\mathbf{r=r}(t)$ and $\mathbf{v=v}(t)=\frac{d}{dt}\mathbf{r}%
(t)$ its position and velocity, $\mathbf{V}\left( \mathbf{r,}t\right) $ the
local fluid velocity and $\mathbf{u}=\mathbf{v-V}(\mathbf{r},t)$ the
relative particle velocity, we intend to prove that $\mathbf{x}%
_{th}(t)=\left\{ \mathbf{r}(t),\mathbf{v}(t)\right\} $, with
\begin{eqnarray}
&&\left. \frac{d}{dt}\mathbf{r}(t)=\mathbf{v}(t),\right.  \label{EQ.0-a} \\
&&\left. \mathbf{v}(t)=\mathbf{V}(\mathbf{r},t)+\mathbf{u}(t),\right.
\label{EQ.0-b}
\end{eqnarray}%
where%
\begin{eqnarray}
\mathbf{u}(t) &=&\mathbf{n}(\mathbf{r},t)u(t),  \label{EQ.1} \\
u(t) &=&v_{th}(\mathbf{r},t)\equiv \sqrt{\frac{2p_{1}(\mathbf{r},t)}{\rho
_{o}}},  \label{EQ.2}
\end{eqnarray}%
is a particular solution of the NS dynamical system. In particular, $v_{th}$
denotes the \textit{thermal velocity} produced by the kinetic pressure $%
p_{1}(\mathbf{r},t)$ [which is, in turn, related to the fluid pressure, see
Eq.(\ref{kinpressure}) below], while $\mathbf{n}(\mathbf{r},t)$ is the
rotating unit vector which satisfies the initial-value problem%
\begin{equation}
\left\{
\begin{array}{c}
\frac{d}{dt}\mathbf{n}(\mathbf{r}t,t)=\mathbf{n}(\mathbf{r},t)\times \mathbf{%
\Omega }(\mathbf{r},t), \\
\mathbf{n}(\mathbf{r}(t_{o}\ ),t_{o})=\mathbf{n}(\mathbf{r}_{o},t_{o}).%
\end{array}%
\right.  \label{EQ.3}
\end{equation}%
Here $\mathbf{\Omega }(\mathbf{r},t)$ is the angular velocity
\begin{equation}
\mathbf{\Omega }(\mathbf{\mathbf{r}},t)\mathbf{=}a\mathbf{\xi ,}
\label{EQ.4}
\end{equation}%
produced only by the \emph{fluid vorticity} $\mathbf{\xi =}\nabla \times
\mathbf{V,}$ while the unit vector $\mathbf{n}(\mathbf{r},t)$ is subject to
the requirement
\begin{equation}
\mathbf{n}(\mathbf{r},t)\cdot \nabla p_{1}(\mathbf{r},t)=0
\label{ORTHOGONALITY of n}
\end{equation}%
(\emph{constraint of orthogonality}). In Eq.(\ref{EQ.4}) $a\equiv a(\mathbf{r%
},t)$ is the real scalar parameter which characterizes the mean-field force $%
\mathbf{F}$ \cite{Tessarotto2007a}. The arbitrariness of $a$ implies that
the definition of the angular velocity $\mathbf{\Omega }(\mathbf{\mathbf{r}}%
,t)$ is non-unique, a feature which can be viewed as as kind of \emph{%
kinematic indeterminacy }of the theory. In Ref.\cite{Tessarotto2007a} $a$
was determined based on phenomenological arguments as
\begin{equation}
a=1/2.  \label{CHOICE of a}
\end{equation}%
Here we intend to show, however, that the requirement of the existence of
the solutions of the type (\ref{EQ.0-a})-(\ref{ORTHOGONALITY of n}) actually
imposes also a \emph{kinematic constraint} on the same parameter, i.e., that
there results also \emph{\ }%
\begin{equation}
\left. a\mathbf{n\cdot \omega }+\mathbf{n\cdot \nabla }\frac{d}{dt}%
p_{1}=0,\right.  \label{kinematic constraint}
\end{equation}%
where $\mathbf{\omega \equiv }\left( \nabla \times \mathbf{V}\right) \times
\mathbf{\nabla }p_{1}.$ In other words (see THM.1) the parameter $a$ can
also be defined in such a way that, if at the initial time $t_{o},$ $\mathbf{%
n}(\mathbf{r}_{o},t_{o})$ satisfies the constraint of orthogonality (with
respect to $\nabla _{o}p_{1}(\mathbf{r}_{o},t_{o}),$ then at any time $t,$ $%
\mathbf{n}(\mathbf{r}(t),t)$ will satisfy the same constraint provided (\ref%
{kinematic constraint}) is satisfied too. Therefore, in such a case the
relative velocity of a generic thermal tracer particle $\mathbf{u}(t)$ is
\emph{a rotating vector orthogonal to }$\nabla p_{1}(\mathbf{r},t).$ In
particular, it is found that while its magnitude $u(t)$ depends, according
to Eq.(\ref{EQ.2}), only on the kinetic pressure $p_{1}(\mathbf{r},t),$ the
angular rotation velocity $\mathbf{\Omega }(\mathbf{r},t)\mathbf{,}$ which
characterizes the direction of the relative velocity, \emph{is produced by
fluid vorticity}.

Main consequence of the theory here presented is that dynamics of an
incompressible NS fluid, i.e., the fluid fields and the complete set of
fluid equations, can be represented solely in terms of the kinetic state of
the thermal tracer particles (see THM.2). The goal is reached by proving
that the dynamics of these particles uniquely generates, via a suitable 1-D
statistics, the time-evolution of the fluid fields.

\subsection{Scheme of presentation}

The general framework (of this work) is provided by the phase-space
representation for an incompressible NS fluid previously developed (see in
particular Tessarotto and coworkers \cite%
{Tessarotto2004,Ellero2005,Tessarotto20083,Tessarotto20081}). In Sec. 2 the
relevant aspect of the theory are recalled, with particular reference to the
so-called \textit{Hopf-Rosen-Edwards} (HRE) \textit{approach} \cite%
{Hopf1952,Rosen1971,Edwards1964} (see also\ Monin and Yaglom, 1975 \cite%
{Monin1975} and Pope, 2000 \cite{Pope2000}) and the IKT approach developed
in Refs.\cite{Tessarotto2004,Ellero2005}. This permits us to identify the NS
dynamical system, via the construction of a suitable mean-field force. After
displaying the corresponding phase-space Lagrangian formulation (in Sec. 3),
the existence of the thermal tracer-particles solutions is proven in THM.1
(see Section 4). Finally the statistics of thermal tracer particles is
investigated in Sec. 5 (THM.2). This is shown to depend only on the
direction of the tracer-particle relative velocity, defined by a coplanar
family of unit vectors.

\section{2 - Phase-space approaches to fluid dynamics}

A fundamental aspect of fluid dynamics is the construction of phase-space
approaches based on \textit{classical statistical mechanics }(CSM). Indeed,
phase-space techniques are well known both in classical and quantum fluid
dynamics. In this connection, a particular viewpoint is represented by the
class of so-called \textit{inverse problems}, involving the search of an
inverse kinetic theory (IKT) \textit{able to yield the complete set of fluid
equations for the fluid fields}, via the introduction of a suitable
phase-space PDF $f(\mathbf{x,}t;Z)$ on the restricted phase-space $\Gamma $\
\cite{Tessarotto2004}$.$ This requires, by assumption, that the PDF must
depend at least functionally, i.e., via appropriate "moments", on the
(deterministic) fluid fields $\left\{ Z\right\} $ characterizing the fluid
(or more generally on a suitable subset of $\left\{ Z\right\} $).

A particular realization for $f(\mathbf{x,}t;Z),$ holding for an
incompressible NS fluid, is provided by the well-known \textit{%
Hopf-Rosen-Edwards} (HRE) \textit{approach} \cite%
{Hopf1952,Rosen1971,Edwards1964} (see also\ Monin and Yaglom, 1975 \cite%
{Monin1975} and Pope, 2000 \cite{Pope2000}). In this case $f(\mathbf{x,}t;Z)$
satisfies the Liouville equation [or \textit{inverse kinetic equation}]
\begin{equation}
L(\mathbf{r,v},t;Z,f)f(\mathbf{x,}t;Z)=0,  \label{Eq---1}
\end{equation}%
with $f(\mathbf{x,}t;Z)\equiv f_{H}(\mathbf{x,}t;Z)$ denoting the PDF
\begin{equation}
f_{H}(\mathbf{x,}t;Z)=\delta \left( \mathbf{v-V}(\mathbf{r,}t)\right)
\label{Eq--1a}
\end{equation}%
and $L(\mathbf{r,v},t;Z,f)$ a suitable Liouville streaming operator. Here
the notation is standard \cite{Ellero2005}. Thus $\mathbf{x}=(\mathbf{r,v})$
is a state-vector which spans the phase space $\Gamma \equiv \Omega \times
U, $ $\Omega $ and $U\equiv
\mathbb{R}
^{3}$ denoting respectively the configuration and velocity spaces. Moreover,
$L(\mathbf{r,v},t;Z,f)$ reads
\begin{eqnarray}
&&\left. L(\mathbf{r,v},t;Z,f)\cdot \equiv \frac{\partial }{\partial t}\cdot
+\frac{\partial }{\partial \mathbf{x}}\cdot \left\{ \mathbf{X}(\mathbf{x}%
,t;Z)\cdot \right\} \equiv \right.  \label{Eq---1b} \\
&&\left. \equiv \frac{\partial }{\partial t}\cdot +\mathbf{v\cdot \nabla }%
\cdot +\frac{\partial }{\partial \mathbf{v}}\cdot \left\{ \mathbf{F}(\mathbf{%
x},t;Z)\cdot \right\} \right.  \notag
\end{eqnarray}%
and $\mathbf{X}(\mathbf{x},t;Z)$ is the vector field
\begin{equation}
\mathbf{X}(\mathbf{x},t;Z)=\left\{ \mathbf{v,F}\right\} .
\label{vector field X}
\end{equation}%
In particular, in the HRE approach $\mathbf{F}$ is identified with $\mathbf{%
F\equiv F}_{H},$ i.e., the total fluid force per unit mass acting on each
fluid element [see Eq.(\ref{2c}) in Appendix A]. It follows by construction
that:

\begin{itemize}
\item the first two velocity moments of $\ f(\mathbf{x,}t;Z)$ are
necessarily
\begin{eqnarray}
\int\limits_{U}d\mathbf{v}f(\mathbf{x,}t;Z) &=&1,  \label{Eq--1cc} \\
\int\limits_{U}d\mathbf{vv}f(\mathbf{x,}t;Z) &=&\mathbf{V}(\mathbf{r,}t);
\label{Eq---1ccc}
\end{eqnarray}

\item the corresponding moment equations obtained above coincide manifestly
with the complete set of fluid equations [see Eqs.(\ref{1b}) and (\ref{1c})];

\item since the PDF $f(\mathbf{x,}t;Z)$ of Eq.(\ref{Eq--1a}), is independent
of the fluid pressure$,$ $p(\mathbf{r,}t)$ cannot, manifestly, be
represented as a \ moment of the same PDF.
\end{itemize}

However, a more general viewpoint is represented by the class of so-called
\textit{complete IKT's}\emph{\ }able to yield as moments of the PDF the
\textit{whole set of \ fluid fields }$\left\{ Z\right\} $ which appear in
the fluid equations\textit{. }A theory of this type has been recently
developed by Tessarotto and coworkers \cite{Tessarotto2004,Ellero2005}.%
\textit{\ }In the case of INSE the theory must include necessarily, among
the PDF moments, also the fluid pressure. This requires (\textit{Principle
of correspondence}) that, besides Eqs.(\ref{Eq--1cc}) and (\ref{Eq---1ccc}),
there must exist a suitable phase-space function $G(\mathbf{x,}t)$ for which
it results identically in $\overline{\Omega }\times I$
\begin{equation}
p(\mathbf{r},t)=\int\limits_{U}d\mathbf{v}f(\mathbf{x,}t;Z)G(\mathbf{x,}t).
\label{Eq--1d}
\end{equation}%
In particular, one can show that a possible choice is provided by the
position \cite{Ellero2005}
\begin{equation}
G(\mathbf{x,}t)=\frac{\rho _{o}u^{2}}{3}-p_{0}(t)-\phi (\mathbf{r},t),
\end{equation}%
where $\mathbf{u}$ is the relative velocity $\mathbf{u=v-V}(\mathbf{\mathbf{r%
},}t)$ and $p_{0}(t)>0$ is a strictly positive smooth real function to be
defined (see below).

The actual construction of a complete IKT of this type relies on classical
statistical mechanics \cite{Tessarotto2006, Tessarotto2007a}, and its
formulation comprises the usual axioms of CSM. It follows that, again, $f(%
\mathbf{x,}t;Z)$ must fulfill necessarily the Liouville equation (\ref%
{Eq---1}) (which can be intended as a \emph{phase-space Eulerian equation}).
Here by construction $\mathbf{X}(\mathbf{x},t;Z)$ is so defined that the
fluid fields $\left\{ \mathbf{V}(\mathbf{r,}t),p(\mathbf{r,}t)\right\} $
must be uniquely determined by the moment equations (\ref{Eq---1ccc}) and (%
\ref{Eq--1d}).\ It is important to stress that, by suitably specifying the
functional class $\left\{ f(\mathbf{x,}t;Z)\right\} $,\ the actual form of
the vector field $\mathbf{X}(\mathbf{x},t;Z)$ can be \textit{explicitly
constructed} \cite{Ellero2005} and proven \textit{to be unique }\cite%
{Tessarotto2007a,Tessarotto2006,Tessarotto2007}. Thus, a particular solution
of Eq.(\ref{Eq---1}) can be obtained by requiring that $f(\mathbf{x,}t;Z)$
for any $\left( \mathbf{x,}t\right) \in \Gamma \times I$ satisfies the two
entropic principles:

\begin{itemize}
\item \textit{Axiom of maximum entropy }(also known as principle of entropy
maximization (PEM); Jaynes, 1957 \cite{Jaynes1957}). This requires that: 1) $%
f(\mathbf{x,}t;Z)$ be a strictly-positive ordinary function which admits for
all $t\in I$ the moment
\begin{equation}
S(f(t))=-\int\limits_{\Gamma }d\mathbf{x}f(\mathbf{x,}t;Z)\ln f(\mathbf{x,}%
t;Z),  \label{A-1}
\end{equation}%
$S(f(t))$ denoting the Boltzmann-Shannon (BS) entropy\textit{\ }associated
to the PDF $f(\mathbf{x,}t;Z)$; 2) that\ for all $t\in I$ the probability
density $f(t)\equiv f(\mathbf{x,}t,Z)$ satisfies the \emph{constrained
maximal variational principle\ }%
\begin{equation}
\delta S(f(t))=0,  \label{A-2}
\end{equation}%
with $f(t)$ required to satisfy the constraints (\ref{Eq--1cc}) and (\ref%
{Eq---1ccc}) and (\ref{Eq--1d});

\item \textit{Axiom of conservation of entropy} (also known as constant
H-theorem \cite{Tessarotto2007a}). This requires that there results
identically for all $t\in I$
\begin{equation}
\frac{\partial }{\partial t}S(f(t))=0.  \label{constant-H}
\end{equation}
\end{itemize}

Imposing PEM requires necessarily that $f(\mathbf{x,}t;Z)$ must coincide
with a local Maxwellian distribution \cite{Ellero2005,Tessarotto2007a}
\begin{equation}
f_{M}(\mathbf{u};p_{1}(\mathbf{r},t))=\frac{1}{\pi ^{2}v_{th}(\mathbf{r},t)}%
\exp \left\{ -\frac{u^{2}}{v_{th}(\mathbf{r},t)}\right\} ,
\label{Maxwellian0}
\end{equation}%
where $\mathbf{u=v-V}(\mathbf{r},t),$
\begin{equation}
v_{th}(\mathbf{r,}t)=\sqrt{2p_{1}(\mathbf{r,t})/\rho _{o}},
\label{thermal velocity}
\end{equation}%
is the \textit{thermal velocity due to the kinetic pressure}
\begin{equation}
p_{1}(\mathbf{r},t)=p_{0}(t)+p(\mathbf{r},t)-\phi (\mathbf{r},t),
\label{kinpressure}
\end{equation}%
and the pseudo-pressure $p_{0}(t_{o})>0$ is uniquely determined in such a
way to assure that in the time interval $I$ the constant H-theorem [Eq.(\ref%
{constant-H})] is satisfied identically. As a further consequence, one can
prove \cite{Tessarotto2007a} that the requirement that
\begin{equation}
f(\mathbf{x,}t;Z)\equiv f_{M}(\mathbf{u};p_{1}(\mathbf{r},t))
\label{Maxwellian case}
\end{equation}%
be a solution of Eq.(\ref{Eq---1}) implies necessarily that the vector field
$\mathbf{F}$ depends functionally on $f_{M}$ and $\left\{ Z\right\} ,$ hence
it can be intended as a mean-field force \cite{Tessarotto2004}. As pointed
out previously \cite{Tessarotto2007a}, the form of $\mathbf{F}$ is actually
non-unique. More precisely it can be cast in the form:%
\begin{equation}
\mathbf{F}(\mathbf{x,}t;f_{M};Z;a)=\mathbf{F}_{0}(\mathbf{x,}t;f_{M};Z;a)+%
\mathbf{F}_{1}(\mathbf{x,}t;f_{M};Z),  \label{F}
\end{equation}%
\begin{equation}
\mathbf{F}_{0}(\mathbf{x,}t;f_{M};a)=\frac{1}{\rho _{0}}\mathbf{f}_{R}+%
\mathbf{D}(\mathbf{r},t;a)\mathbf{+}\nu \nabla ^{2}\mathbf{V,}  \label{F-2}
\end{equation}%
\begin{equation}
\mathbf{F}_{1}(\mathbf{x,}t;f_{M})=\frac{1}{2}\mathbf{u}A\mathbf{(r,}%
t;f)+\Delta \mathbf{F}_{1}(\mathbf{x,}t;f_{M}),  \label{F-3}
\end{equation}%
\begin{equation}
\Delta \mathbf{F}_{1}(\mathbf{x,}t;f_{M})=\frac{v_{th}}{2p_{1}}\nabla
p_{1}\left\{ \frac{u^{2}}{v_{th}}-\frac{3}{2}\right\} ,  \label{F-4}
\end{equation}%
where $\mathbf{D}(\mathbf{r},t;a)$ is the convective term%
\begin{equation}
\mathbf{D}(\mathbf{r},t;a)=\mathbf{u}\cdot \nabla \mathbf{V}b+a\nabla
\mathbf{V\cdot \mathbf{u,}}  \label{convective
term}
\end{equation}%
$b=1-a,$ and $a\equiv a(\mathbf{r},t)$ is an arbitrary scalar real function.
Finally, the vector field $A(\mathbf{r,}t;f_{M})$ is now defined as\emph{\ }%
\begin{eqnarray}
&&\left. A(\mathbf{r,}t;f_{M})\equiv \frac{1}{p_{1}}\frac{\partial }{%
\partial t}p_{1}-\right.  \label{D-p_1} \\
&&\left. -\frac{\rho _{o}}{p_{1}}\left[ \frac{\partial }{\partial t}%
V^{2}/2+\nabla \cdot \left( \mathbf{V}V^{2}/2\right) -\frac{1}{\rho _{o}}%
\mathbf{V\cdot f}-\nu \mathbf{V\cdot }\nabla ^{2}\mathbf{V}\right] \equiv
\right.  \notag \\
&\equiv &\frac{1}{p_{1}}\left[ \frac{D}{Dt}p_{1}+\frac{\rho _{o}V^{2}}{2}%
\nabla \cdot \mathbf{V}\right] ,  \notag
\end{eqnarray}%
where a term proportional to $\nabla \cdot \mathbf{V}$ [and hence
identically vanishing when the isochoricity condition is satisfied; see Eq.(%
\ref{1b}) in Appendix A] has been added.

Here we stress that:

\begin{enumerate}
\item as previously pointed out \cite{Tessarotto2007} the convective term $%
\mathbf{D}(\mathbf{r},t;a)$ is non-unique due to the indeterminacy of the
parameter $a.$ This implies also the non-uniqueness of the NS dynamical
system (see next Section). The choice (\ref{CHOICE of a}) for the parameter $%
a$ then corresponds to the symmetry condition \cite{Tessarotto2007}%
\begin{equation}
a=b=1/2;
\end{equation}

\item the complete IKT holds in principle for arbitrary choices of $a,$ to
be considered either a real constant or a smooth real function $a(\mathbf{r}%
,t);$

\item as indicated elsewhere \cite{Ellero2005} the present theory can also
be generalized in a straightforward way also to non-Maxwellian initial
distribution functions $f(\mathbf{x,}t;Z\mathbb{)}$.
\end{enumerate}

\section{3 - Phase-space Lagrangian formulation}

The results of the previous Section permit us to formulate in a
straightforward way the phase-space Lagrangian formulation for an
incompressible NS fluid. The complete IKT implies, in fact, that there
exists a phase-space classical dynamical system (called \textit{NS}\emph{\ }%
\textit{dynamical system})%
\begin{equation}
\mathbf{x}_{o}\rightarrow \mathbf{x}(t)\equiv \left\{ \mathbf{r}(t),\mathbf{v%
}(t)\right\} =T_{t,t_{o}}\mathbf{x}_{o}\equiv \chi (\mathbf{x}_{o},t_{o},t),
\label{classical dynamical system}
\end{equation}%
$T_{t,t_{o}}$ denoting the corresponding evolution operator, which generates
the time-evolution of the complete set of fluid fields, i.e., is such that%
\begin{equation}
\left\{ Z((\mathbf{r}(t),t)\right\} =T_{t,t_{o}}\left\{ Z_{o}(\mathbf{r}%
_{o})\right\}  \label{time evolution}
\end{equation}%
[here $Z_{o}(\mathbf{r}_{o})\equiv Z(\mathbf{r}_{o},t_{o})$ denotes the
initial fluid fields evaluates at $t=t_{o};$ see Eq.(\ref{1d})]. This means
that the NS dynamical system is identified with the flow generated by the
vector field $\mathbf{X}$ defined by Eq.(\ref{vector field X}).\ In view of
Eqs.(\ref{F})-(\ref{D-p_1}) this must be considered of the form%
\begin{equation}
\mathbf{X=X}(\mathbf{x,}t;f;Z;a).  \label{vector field X-1}
\end{equation}%
Hence,\textit{\ }$\mathbf{x}(t)$ is obtained as the solution of the
initial-value problem defined by Eq.(\ref{EQ.0}), where, in particular, the
mean-field force $\mathbf{F}$\textit{\ }is expected to differ generally from%
\textit{\ }$\mathbf{F}_{H}$ as defined by Eq.(\ref{2c}) [see Appendix].
Moreover the state-vector $\mathbf{x}=(\mathbf{r,v})$ spans by assumption
the phase space $\Gamma \equiv \Omega \times U,$ while $\Omega $ and $U$
denote appropriate configuration and velocity spaces. \emph{The integral
curves} $\mathbf{x}(t)$ \emph{of} \emph{Eq.(\ref{EQ.0})\ can therefore be
interpreted as Lagrangian phase-space trajectories of tracer particles,
whose dynamics is determined uniquely by the time evolution of the fluid}.
The latter can also be cast in an equivalent \emph{phase-space Lagrangian
form}. By assumption the initial-value problem (\ref{EQ.0}) is well-posed,
namely it defines a suitably smooth diffeomorphism. Here, by construction:

\begin{itemize}
\item denoting by $\mathbf{x}(t)=\chi (\mathbf{x}_{o},t_{o},t)$ the solution
of the initial-value problem (\ref{EQ.0}), $\mathbf{x}_{o}=\chi (\mathbf{x}%
(t),t,t_{o})$ is its inverse. Both are assumed to be suitably smooth
functions of the relevant parameters;

\item both $\mathbf{x}(t)=\chi (\mathbf{x}_{o},t_{o},t)$ and $\mathbf{x}%
_{o}=\chi (\mathbf{x}(t),t,t_{o})$ \ identify admissible LP's of the
dynamical system;

\item $\mathbf{r}(t)$ is the Lagrangian trajectory which belongs to the
fluid domain $\Omega $;

\item $\mathbf{v}(t)$ and\ $\mathbf{F}(\mathbf{r}(t),t;f)$ are respectively
the Lagrangian velocity and acceleration, both spanning the vector space $%
\mathbb{R}
^{3}.$ In particular, \ $\mathbf{F}(\mathbf{r}(t),t;f),$ which is defined by
Eqs.(\ref{F})-(\ref{D-p_1}), and depends functionally on the kinetic
probability density $f(\mathbf{x},t),$ is the \emph{Lagrangian acceleration
which corresponds to an arbitrary kinetic probability density} $f(\mathbf{x}%
,t);$

\item $f(\mathbf{x},t)$ is a particular solution of the inverse kinetic
equation (\ref{Eq---1}).
\end{itemize}

It follows that the Jacobian [$J(\mathbf{x}(t),t)=\left\vert \frac{\partial
\mathbf{x}(t)}{\partial \mathbf{x}_{o}}\right\vert $] of the map $\mathbf{x}%
_{o}\rightarrow \mathbf{x}(t),$ which is generated by Eq.(\ref{EQ.0}) for $%
f\equiv f_{M}(\mathbf{x},t)$ reads
\begin{eqnarray}
&&\left. J(\mathbf{x}(t),t)\equiv \left\vert \frac{\partial \mathbf{x}(t)}{%
\partial \mathbf{x}_{o}}\right\vert =\right. \\
&=&\frac{v_{th}(t)}{v_{th}(t_{o})}\exp \left\{ -\frac{u^{2}(t_{o})}{%
v_{th}(t_{o})}+\frac{u^{2}(t)}{v_{th}(t)}\right\} .  \notag
\end{eqnarray}

Furthermore, it is immediate to prove that the inverse kinetic equation can
be written in Lagrangian form. This yields the so-called \textit{integral
Liouville equation}

\begin{equation}
J(\mathbf{x}(t),t)f(\mathbf{x}(t),t;Z)=f(\mathbf{x}_{o},t_{o},Z),
\label{Lagrangian IKE}
\end{equation}%
with $f(\mathbf{x}_{o},t_{o},Z)$ denoting \ the initial PDF and $J(\mathbf{x}%
(t),t)=\left\vert \frac{\partial \mathbf{x}(t)}{\partial \mathbf{x}_{o}}%
\right\vert $ the Jacobian of the map (\ref{classical dynamical system}). As
a consequence, for all $\left( \mathbf{x},t\right) \in \overline{\Gamma }%
\times I$ (with $\overline{\Gamma }\equiv \overline{\Omega }\times U$) the
PDF $f(\mathbf{x},t;Z)$ can be represented explicitly in terms of the
initial PDF and reads
\begin{equation}
f(\mathbf{x},t;Z)=\frac{1}{J(t;Z)}f(\chi (\mathbf{x},t,t),t_{o},Z),
\label{Lagrangian-IKE-2}
\end{equation}%
where $J(t;Z)\equiv J(\mathbf{x},t),$ $\chi (\mathbf{x},t,t)\equiv
T_{t,t_{o}}^{-1}\mathbf{x}$ and $\mathbf{x}\equiv \mathbf{x}(t).$ In
particular, in terms of Eq.(\ref{Lagrangian-IKE-2}) the fluid fields can be
explicitly evaluated, yielding:%
\begin{eqnarray}
\mathbf{V}(\mathbf{r,}t) &=&\int\limits_{U}d\mathbf{vv}\frac{1}{J(t;Z)}%
f(\chi (\mathbf{x},t,t),t_{o},Z), \\
p(\mathbf{r},t) &=&\int\limits_{U}d\mathbf{v}\frac{1}{J(t;Z)}f(\chi (\mathbf{%
x},t,t),t_{o},Z) \\
&&\left[ \frac{\rho _{o}u^{2}}{3}-p_{0}(t)-\phi (\mathbf{r},t)\right] .
\notag
\end{eqnarray}%
From the \textit{mathematical standpoint} main consequences of the IKT
representation are that: 1) the Lagrangian formulation (of IKT) is uniquely
specified by the proper definition of a suitable family of phase-space LP's;
2) Eq.(\ref{Lagrangian-IKE-2}) uniquely specifies the time-evolution of the
Eulerian PDF, $f(\mathbf{x}(t),t),$ which is represented in terms of the
initial distribution function $f_{o}(\mathbf{x}_{o})$ and the LP's defined
by the INSE dynamical system; 3) the time-evolution of the fluid fields $%
\left\{ \rho =\rho _{o}>0,\mathbf{V},p\geq 0,T>0,S_{T}\right\} $ is uniquely
specified via the PDF $f(\mathbf{x}(t),t);$ 4) Eq.(\ref{Lagrangian-IKE-2})
also provides the connection between Lagrangian and Eulerian viewpoints. In
fact the Eulerian PDF, $f(\mathbf{x},t),$ is simply obtained from Eq.(\ref%
{Lagrangian-IKE-2}) by letting $\mathbf{x}=\mathbf{x}(t)$ in the same
equation. As a result, the Eulerian and Lagrangian formulations of IKT, and
hence of the underlying moment (i.e., fluid) equations, are manifestly
equivalent$.$ From the \textit{physical viewpoint}, it is worth mentioning
that the LP's here defined can be interpreted as phase-space trajectories of
the particles of the fluid, to be considered as a set of "classical
molecules", i.e., point particles with prescribed mass, which interact with
the fluid only via the action of a suitable mean-field force kind. The
ensemble motion of these particles has been defined in such a way that it
uniquely determines the time evolution \textit{both} of the kinetic
distribution functions \textit{and} of the relevant fluid fields which
characterize the NS fluid.

\section{4 - The dynamics of thermal tracer particles}

An interesting issue is the search of \emph{possible exact particular
solutions} of the Lagrangian equations (\ref{EQ.0}). The general solution of
Eq.(\ref{EQ.0}) implies
\begin{equation}
\mathbf{v}(t)=\mathbf{V}(\mathbf{r}(t),t)+\mathbf{n}(\mathbf{r}(t),t)u(%
\mathbf{r}(t),t),  \label{THERMAL TRACER DYNAMICS}
\end{equation}%
where $\mathbf{n}(\mathbf{r}(t),t)$ is the rotating unit vector which
satisfies the initial-value problem Eq.(\ref{EQ.3}). For arbitrary
tracer-particle solutions it is obvious that $\mathbf{n}(\mathbf{r}(t),t)$
is an arbitrary unit vector belonging to the unit sphere [$S(U)$] of
velocity space $U\equiv
\mathbb{R}
^{3}.$ Hence, by definition, $S(U)$ is a set of dimension 2$.$ In fact, Eq.(%
\ref{EQ.0}) yields for the relative velocity $\mathbf{u=v-V}(\mathbf{r},t)$
the equation%
\begin{eqnarray}
&&\left. \frac{d}{dt}\mathbf{u}=\mathbf{F}_{u}(\mathbf{x,}%
t;f_{M};Z;a),\right.  \label{pr} \\
&&\mathbf{F}_{u}(\mathbf{x,}t;f_{M};Z;a)=a\mathbb{\nabla }\mathbf{V\cdot u}-a%
\mathbf{\mathbf{u}\cdot \nabla \mathbf{V}+}  \label{prrr} \\
&&+\frac{\mathbf{u}}{2p_{1}}\left[ \frac{D}{Dt}p_{1}+\frac{\rho _{o}V^{2}}{%
2p_{1}}\nabla \cdot \mathbf{V}\right] +  \notag \\
&&+\frac{v_{th}}{2p_{1}}\mathbf{\nabla }p_{1}\left\{ \frac{u^{2}}{v_{th}}-%
\frac{1}{2}\right\} .  \notag
\end{eqnarray}%
It follows
\begin{equation}
\frac{d}{dt}\frac{u^{2}}{2}\mathbf{=}\frac{u^{2}}{2}\frac{1}{p_{1}}\left[
\frac{d}{dt}p_{1}+\frac{\rho _{o}V^{2}}{2}\nabla \cdot \mathbf{V}\right] -%
\frac{1}{2\rho _{o}}\mathbf{u\cdot \nabla }p_{1}.  \label{prev2}
\end{equation}%
Furthermore, Eq.(\ref{pr}), in view of Eq.(\ref{prev2}), implies for $%
\mathbf{n}$
\begin{eqnarray}
&&\left. \frac{d}{dt}\mathbf{n}(\mathbf{r}t,t)=\mathbf{n}(\mathbf{r}%
,t)\times \mathbf{\Omega }(\mathbf{r},t)-\right.  \label{EQ.for n} \\
&&-\frac{1}{\rho _{o}}\left\{ \frac{u^{2}}{v_{th}}-\frac{1}{2}\right\}
\mathbf{n}(\mathbf{r},t)\times \left[ \mathbf{n}(\mathbf{r},t)\times \mathbf{%
\nabla }p_{1}\right] .  \notag
\end{eqnarray}%
There it follows:

\textbf{Theorem 1 - Existence of thermal tracer particles}

\emph{If the fluid fields }$\left\{ \mathbf{V}(\mathbf{r},t),p(\mathbf{r}%
,t)\right\} $ \emph{are classical solutions of INSE [see Eqs.(\ref{1a})-(\ref%
{1e}) in Appendix A], then a particular solution }$\mathbf{x}(t)\equiv $ $%
\mathbf{x}_{Th}(t)$ \emph{of Eq.(\ref{EQ.0}) is provided by Eqs.(\ref{EQ.0-a}%
)-(\ref{EQ.4}), subject to the orthogonality and kinematic constraints (\ref%
{ORTHOGONALITY of n}) and (\ref{kinematic constraint}).}

\textit{PROOF}

First we notice, invoking Eq.(\ref{prev2}) and assuming that $\mathbf{V}(%
\mathbf{r},t)$ satisfies in particular the isochoricity condition [Eq.(\ref%
{1c}), in Appendix A], that Eq.(\ref{EQ.2}) implies necessarily
\begin{equation}
\frac{d}{dt}\frac{u^{2}}{2}\mathbf{=}\frac{u^{2}}{2}\frac{1}{p_{1}}\frac{d}{%
dt}p_{1}.  \label{EQUATION}
\end{equation}

It follows that the unit vector $\mathbf{n}\left( \mathbf{r},t\right) $ must
satisfy identically Eq.(\ref{ORTHOGONALITY of n}). On the other hand, thanks
to Eq.(\ref{EQ.3}), according to Eq.(\ref{EQ.for n}), $\mathbf{n}\left(
\mathbf{r},t\right) $\ is a rotating unit vector with angular velocity $%
\mathbf{\Omega }(\mathbf{r},t)$ defined by Eq.(\ref{EQ.4}). Hence, the
condition of orthogonality Eq.(\ref{ORTHOGONALITY of n}) requires
\begin{equation}
\frac{d}{dt}\mathbf{n\cdot \nabla }p_{1}+\mathbf{n\cdot }\frac{d}{dt}\mathbf{%
\nabla }p_{1}=0  \label{START-2}
\end{equation}%
too. Eq.(\ref{EQ.3}) then implies manifestly also Eq.(\ref{kinematic
constraint}),\emph{\ }which can be viewed as a kinematic constraint for the
parameter $a,$ to be considered as dynamical function. If $\mathbf{\omega }$
is non-vanishing and the scalar product $\mathbf{n\cdot \omega }$ is defined
so that it is $\neq 0$ it is obvious that the previous equation can always
be satisfied. On the contrary, if locally $\mathbf{\omega =0,}$\ this can
happen either if 1) $\mathbf{\xi =0,}$ 2) $\mathbf{\nabla }p_{1}=0$ or 3) $%
\mathbf{\xi \parallel \nabla }p_{1}.$

In the first case, it follows Eq.(\ref{EQ.3}) reduces to $\frac{d}{dt}%
\mathbf{n=0}$ which requires also $\mathbf{n\cdot \nabla }p_{1}=0$ and
moreover $\mathbf{V}$ is potential, i.e., there exists a real scalar $G(%
\mathbf{r,}t)$ such that $\mathbf{V=\nabla }G(\mathbf{\mathbf{r,}}t).$ This
implies that $G(\mathbf{\mathbf{r,}}t)$ must satisfy the PDE%
\begin{equation}
\frac{\partial }{\partial t}\mathbf{\nabla }G+\frac{1}{2}\mathbf{\nabla }%
\left( \left\vert \nabla G\right\vert ^{2}\right) +\mathbf{\nabla }p_{1}=0,
\end{equation}%
namely locally it must be
\begin{equation}
\mathbf{n\cdot }\left[ \frac{\partial }{\partial t}\mathbf{\nabla }G+\frac{1%
}{2}\mathbf{\nabla }\left( \left\vert \nabla G\right\vert ^{2}\right) \right]
=0
\end{equation}%
too. In the second case the direction of $\mathbf{n}$ is arbitrary so that
it can always be chosen in such a way to satisfy Eq.(\ref{EQ.3}). Finally,
in the third case if $\mathbf{n}$ is orthogonal $\mathbf{\nabla }p_{1}$ it
remains always so by construction thanks to Eq.(\ref{EQ.3}). Let us now
impose the constraints (\ref{ORTHOGONALITY of n}), (\ref{kinematic
constraint}) and consider an arbitrary initial condition of the form
\begin{equation}
\mathbf{x}(t_{o})=\left\{ \mathbf{r}(t_{o})=\mathbf{r}_{o},\mathbf{v}(t_{o})=%
\mathbf{v}_{o}\right\} ,  \label{initial cond-1}
\end{equation}%
with%
\begin{eqnarray}
&&\left. \mathbf{v}_{o}=\mathbf{V}(\mathbf{r}_{o},t_{o})+\mathbf{n}(\mathbf{r%
}_{o},t_{o})u(t_{o}),\right.  \label{initial cond-2} \\
&&\left. u(t_{o})=v_{th}(\mathbf{r}_{o},t_{o})\equiv \sqrt{\frac{2p_{1}(%
\mathbf{r}_{o},t_{o})}{\rho _{o}}},\right.  \label{initial cond-3} \\
&&\left. \mathbf{n}(\mathbf{r}_{o},t_{o})\cdot \nabla _{o}p_{1}(\mathbf{r}%
_{o},t_{o})=0\right.  \label{initial cond-4}
\end{eqnarray}%
Here $\left\{ \mathbf{V}(\mathbf{r}_{o},t_{o}),p(\mathbf{r}%
_{o},t_{o})\right\} $ are arbitrary initial fluid fields consistent with
INSE (see Appendix A), while $p_{1}(\mathbf{r}_{o},t_{o})$ and $\nabla
_{o}p_{1}(\mathbf{r}_{o},t_{o})$ are defined according to Eqs.(\ref%
{kinpressure}). Finally, $\mathbf{n}(\mathbf{r}_{o},t_{o})\equiv \mathbf{n}%
_{o}$ is an arbitrary unit vector orthogonal to $\nabla _{o}p_{1}(\mathbf{r}%
_{o},t_{o}).$ It is obvious that the particular solution\ $\mathbf{x}(t)$
which satisfies the previous initial conditions (\ref{initial cond-1})-(\ref%
{initial cond-4}) coincides necessarily with $\mathbf{x}_{TT}(t)$ defined by
Eqs.(\ref{EQ.0-a})-(\ref{EQ.4}). Q.E.D.\

Let us analyze the physical interpretation of the theorem.

A basic consequence is that an arbitrary tracer particle, subject to the
kinematic constraint (\ref{kinematic constraint}) and fulfilling the initial
conditions (\ref{initial cond-1})-(\ref{initial cond-4}) is necessarily a
thermal tracer particle for all $\in I.$ \ Hence, at time $t$\ its relative
velocity is
\begin{equation}
\mathbf{u}(\mathbf{r}(t\ ),t)=\mathbf{n}(\mathbf{r}(t\ ),t)\sqrt{\frac{%
2p_{1}(\mathbf{r}(t\ ),t)}{\rho _{o}},}
\end{equation}%
with $\mathbf{n}$ denoting a rotating unit vector in accordance to Eqs.(\ref%
{EQ.3}) and (\ref{EQ.4}). The consequence is that tracer particles
with initial velocity (\ref{initial cond-1})-(\ref{initial
cond-4}) behave as \textit{thermal particles},\ i.e., at any time
$t$\ their relative velocity [with respect to the local fluid
element] is solely determined by the kinetic pressure
$p_{1}(\mathbf{r},t)$ [in turn related via Eq.(\ref
{kinpressure}) to the fluid pressure]. The direction [$\mathbf{n}(\mathbf{r}%
,t)$] of their relative velocity $\mathbf{u}(\mathbf{r},t)$ depends,
instead, on the angular velocity $\mathbf{\Omega }(\mathbf{\mathbf{r}},t)$
defined by Eq.(\ref{EQ.4}) and depending only on the fluid vorticity.
However, in order to assure the condition of orthogonality (\ref%
{ORTHOGONALITY of n}) for the relative velocity, the kinematic constraint (%
\ref{kinematic constraint}) must be satisfies so that both $\mathbf{\Omega }(%
\mathbf{\mathbf{r}},t)$ and the mean-field force $\mathbf{F}$ are both
uniquely determined. \ This implies that $\mathbf{u}$ necessarily belongs to
the 1-dimensional subset of velocity space $U_{th}\subset U$ \ for which the
orthogonality condition Eq.(\ref{ORTHOGONALITY of n}) holds.

\section{5 - Thermal tracer particle 1-D statistics}

Apart the physical insight afforded by the qualitative behavior of
tracer-particle dynamics, the existence of the particular solution defined
by Eqs.(\ref{EQ.0-a})-(\ref{EQ.4}) is important in order to establish a
phase-space statistical description for incompressible NS fluids. \ In this
section we intend to prove that the dynamics of thermal tracer particles is
\textit{actually sufficient to determine uniquely the time evolution of the
fluid fields}. In other words, to characterize completely the statistical
behavior of the fluid it is sufficient, in principle, to determine \textit{%
only} the dynamics of these tracer particles.

In view of the general form of the solution for thermal tracer particles
[see Eqs.(\ref{EQ.0-a})-(\ref{EQ.4})], it is obvious that the statistics
depends only on the unit vector\ $\mathbf{n}(\mathbf{r},t)$ which defines
the direction of the particle relative velocity with respect to the fluid
[see Eq.(\ref{EQ.3})]. Thanks to the orthogonality constraint (\ref%
{ORTHOGONALITY of n}) placed on the dynamics of thermal tracer particles it
follows that $\mathbf{n}(\mathbf{r},t)$ is orthogonal to $\nabla p_{1}(%
\mathbf{r},t)\equiv \widehat{e}_{z}\left\vert \nabla p_{1}(\mathbf{r}%
,t)\right\vert ,$ i.e., $\mathbf{n}(\mathbf{r},t)$ must belong to a family
of coplanar unit vectors which are tangent to the isobaric surface $\left\{
p_{1}(\mathbf{r},t)=const.\right\} .$ Hence, the only admissible
velocity-space statistics is necessarily one-dimensional (since it is
defined on the subset of velocity space $U_{th}$ define above). This is
obtained by considering $\mathbf{n}(\mathbf{r},t)$ as a \emph{stochastic}
unit vector characterized by a suitable stochastic PDF\emph{\ }$g.$\emph{\ }%
It is obvious that\emph{\ }$g$ is necessarily related to the PDF introduced
in the complete IKT described in the previous section, i.e., to $f(\mathbf{x}%
,t;Z).$ In case (\ref{Maxwellian case}), by parametrizing $\mathbf{n}$ in
cylindrical coordinates
\begin{equation}
\mathbf{n}=\mathbf{n}\left( \mathbf{r,}\vartheta ,t\right) =\widehat{e}%
_{x}\cos \vartheta +\widehat{e}_{y}\sin \vartheta .
\label{parametrization-1D}
\end{equation}%
it follows that $g$ must be necessarily identified with the uniform
distribution
\begin{equation}
g=1/2\pi .  \label{stochastic PDFa}
\end{equation}%
Thus, introducing the stochastic average%
\begin{equation}
\left\langle {}\right\rangle =\int_{0}^{2\pi }d\vartheta g  \label{average-a}
\end{equation}%
[where the angle-integration is performed at constant $\mathbf{r\equiv
\mathbf{r}}(t\ )$ and $t$] it follows by definition that
\begin{eqnarray}
\left\langle 1\right\rangle &=&1,  \label{AV-1a} \\
\left\langle \mathbf{n}\right\rangle &=&\mathbf{0},  \label{AV-1b} \\
\left\langle \mathbf{nn}\right\rangle &=&\frac{1}{2}\underline{\underline{%
\mathbf{I}}}_{2}\equiv \frac{1}{2}\left[ \widehat{e}_{x}\widehat{e}_{x}+%
\widehat{e}_{y}\widehat{e}_{y}\right] .  \label{AV-2a}
\end{eqnarray}

As a consequence it is immediate to show that:

\begin{itemize}
\item the fluid fields $\left\{ \mathbf{V}(\mathbf{r},t),p(\mathbf{r}%
,t)\right\} $ can be related to suitable stochastic averages of thermal
tracer particle dynamics;

\item similarly, by taking the stochastic average of the Lagrangian
equations (\ref{EQ.0}) and (\ref{prev2}), the stochastic-averaged equations
\begin{eqnarray}
\frac{d}{dt}\left\langle \mathbf{v}(t)\right\rangle &=&\left\langle \mathbf{%
F(x,}t;f_{M},Z)\right\rangle ,  \label{stoch-1} \\
\frac{1}{2}\frac{d}{dt}\left\langle u^{2}(t)\right\rangle &=&\left\langle
\mathbf{u}(t)\mathbf{\cdot F}_{u}\mathbf{(x,}t;f_{M},Z)\right\rangle
\label{stoch-2}
\end{eqnarray}%
can be evaluated explicitly for thermal tracer particles.
\end{itemize}

The following result holds.

\textbf{Theorem 2 - 1-D statistics of thermal tracer particles}

\emph{In validity of THM.1, let us assume that} $\mathbf{x}(t)=\left\{
\mathbf{r}(t),\mathbf{v}(t)\right\} $ \emph{is an arbitrary phase-space
trajectory of a thermal tracer particle [i.e., a solution of the form given
by Eqs.(\ref{EQ.0-a})-(\ref{EQ.4})]. Then, due to (\ref{stochastic PDFa})
and (\ref{average-a}), it follows that:}

\emph{1) the stochastic averages }$\left\langle \mathbf{v}(t)\right\rangle $
\emph{and} $\left\langle u(t)^{2}\right\rangle $ \emph{are related to the
local values of the NS fluid fields }$\left\{ \mathbf{V}(\mathbf{r},t),p(%
\mathbf{r},t)\right\} ,$ \emph{namely it results} \emph{\ }
\begin{eqnarray}
\left\langle \mathbf{v}(t)\right\rangle &=&\mathbf{V}(\mathbf{r},t),
\label{AAA-1} \\
\frac{\rho _{o}}{2}\left\langle u(t)^{2}\right\rangle -p_{0}(t)+\phi (%
\mathbf{r},t) &=&p(\mathbf{r},t),  \label{AAA-2}
\end{eqnarray}%
\emph{where }$p_{0}(t)$\emph{\ and }$\phi (\mathbf{r},t)$\emph{\ are defined
by (\ref{kinpressure}),(\ref{2d}) [see Appendix A] and the constant
H-theorem (\ref{constant-H});}

\emph{2) equations (\ref{stoch-1}) and (\ref{stoch-2}) deliver respectively
the NS equation and the isochoricity condition [see Eqs.(\ref{1c}) and Eq.(%
\ref{1b}) in Appendix A].}

\textit{PROOF}

The proof of Eqs.(\ref{AAA-1}) and (\ref{AAA-2}) follows immediately from
Eqs.(\ref{AV-1a})-(\ref{AV-2a}) \ and Eqs.(\ref{EQ.1})-(\ref{EQ.3}).

To evaluate the stochastic average of Eq.(\ref{stoch-1}) we first recall the
identity
\begin{equation}
\frac{d}{dt}\mathbf{v}\mathbf{=}\frac{\partial }{\partial t}\mathbf{V}%
+\left( \mathbf{V+n}u\right) \mathbf{\cdot \nabla V}+\frac{d}{dt}\mathbf{u}.
\end{equation}%
Invoking Eqs.(\ref{pr}) and (\ref{prrr}) it follows%
\begin{eqnarray}
\left\langle \frac{d}{dt}\mathbf{v}\right\rangle &=&\frac{\partial }{%
\partial t}\mathbf{V}(\mathbf{r},t)+\mathbf{V\cdot \nabla V}+
\label{averaged eq} \\
+\left\langle \frac{d}{dt}\mathbf{u}\right\rangle &=&\mathbf{F}%
_{H}+\left\langle \mathbf{F}_{u}(\mathbf{x,}t;f_{M};Z)\right\rangle ,  \notag
\end{eqnarray}%
where $\mathbf{F}_{H}$ is given by Eq.(\ref{2c}) [see Appendix A] and%
\begin{equation}
\left\langle \mathbf{F}_{u}(\mathbf{x,}t;f_{M};Z)\right\rangle =\frac{v_{th}%
}{2p_{1}}\mathbf{\nabla }p_{1}\left\{ \frac{u^{2}}{v_{th}}-\frac{1}{2}%
\right\} =0.
\end{equation}%
Hence Eq.(\ref{averaged eq}) [and Eq.(\ref{stoch-1})] coincides manifestly
with the NS equation (\ref{1c}).

In a similar way it is immediate to evaluate Eq.(\ref{stoch-2}). From Eqs.(%
\ref{EQ.1}) and (\ref{EQ.2}), in validity of the constraint (\ref%
{ORTHOGONALITY of n}), one obtains in fact:%
\begin{equation}
\left\langle \frac{d}{dt}\frac{u^{2}}{2}\right\rangle =\left\langle \mathbf{%
u\cdot F}_{u}(\mathbf{x,}t;f_{M};Z)\right\rangle ,  \label{stoch-6}
\end{equation}%
while under the same conditions from Eq. (\ref{prrr}) it follows%
\begin{eqnarray*}
\left\langle \mathbf{u\cdot }\frac{d}{dt}\mathbf{v}\right\rangle &\mathbf{=}%
&\left\langle \mathbf{u\cdot }\frac{\partial }{\partial t}\mathbf{V}%
\right\rangle +\left\langle \left( \mathbf{V+u}\right) \mathbf{\cdot \nabla
V\cdot u}\right\rangle +\left\langle \frac{d}{dt}\frac{u^{2}}{2}%
\right\rangle = \\
&=&\frac{1}{2}\frac{2p_{1}}{\rho _{o}}\mathbf{\nabla \cdot V+}\frac{D}{Dt}%
\frac{p_{1}}{\rho _{o}}
\end{eqnarray*}%
\begin{equation*}
\left\langle \frac{d}{dt}\frac{u^{2}}{2}\right\rangle =\frac{D}{Dt}\frac{%
p_{1}(\mathbf{r,}t)}{\rho _{o}}
\end{equation*}

\begin{equation}
\left\langle \mathbf{u\cdot F}_{u}(\mathbf{x,}t;f_{M};Z)\right\rangle =\frac{%
1}{\rho _{o}}\left[ \frac{D}{Dt}p_{1}+\frac{\rho _{o}V^{2}}{2}\nabla \cdot
\mathbf{V}\right] .
\end{equation}%
Eqs.(\ref{stoch-2}) and (\ref{stoch-6}) necessarily yield the identity%
\begin{equation}
\frac{V^{2}}{2}\nabla \cdot \mathbf{V}=0.
\end{equation}%
Hence, provided $V^{2}\neq 0,$ also the isochoricity condition [see Eq.(\ref%
{1b}) in Appendix A] is identically fulfilled. Q.E.D.\bigskip

\section{6 - Concluding remarks}

In this paper properties of the NS dynamical system, advancing in time the
state of an incompressible NS fluid, have been investigated. This refers, in
particular, to the proof of the existence of a particular subset of
phase-space trajectories (solutions of the NS dynamical) representing the
dynamics of so-called \emph{thermal tracer particles}. The result has been
reached by imposing suitable constraints to the direction of $\mathbf{u}$
(relative velocity) of tracer particles and on the parameter $a$ which
characterizes its time evolution. In turn, the same constraint determines
uniquely also the mean-field force $\mathbf{F}$ which defines the NS
dynamical system.\ As a main consequence, we have proven (THM.1) that there
exists a family of particular solutions $\mathbf{x}(t)=\left\{ \mathbf{r}(t),%
\mathbf{v}(t)\right\} $ which have the following features:

\begin{enumerate}
\item the magnitude of the relative velocity with respect to the fluid, $%
\mathbf{u}=\mathbf{n}u,$ is determined only by the kinetic pressure $p_{1}$%
\textit{\ }(i.e.,\ tracer particles of this type behave as thermal
particles);

\item the direction of the relative velocity $\mathbf{n}$ is a rotating unit
vector. Its angular velocity is due to the fluid vorticity [see Eq.(\ref%
{EQ.4})];

\item the dimension of the subset of the velocity space $U_{th}\subseteq U$
spanned by these solutions is $\dim (U_{th})=1;$ hence their phase space $%
\Gamma _{th}$=$\Omega \times U_{th}$ is of dimension $\dim (\Gamma _{th})=4.$
\end{enumerate}

Furthermore, the 1-D velocity-space statistics of thermal tracer
particles has been investigated (Section 5). As a main consequence
(see THM.2) :

\begin{enumerate}
\item The state of thermal tracer particles has been proven to determine
uniquely, via a suitable statistics [namely by means of the stochastic
averages (\ref{AAA-1}) and (\ref{AAA-2})], the complete set of fluid fields $%
\left\{ \mathbf{V},p\right\} .$ As a consequence the state of the fluid has
been proven to be uniquely related to that of the thermal tracer particles.

\item The complete set of fluid equations has been represented in terms of
stochastic averages of the phase-space Lagrangian equations for thermal
tracer particles (\ref{stoch-1}) and (\ref{stoch-2}).
\end{enumerate}

These conclusions show that a\emph{\ phase-space Lagrangian dynamics for an
incompressible NS fluid based exclusively on the dynamics of thermal tracer
particles is possible}. The result appears relevant because of its
outstanding physical interpretation. The phase-space Lagrangian dynamics
here determined permits, in fact, to advance in time self-consistently the
fluid fields, i.e., in such a way that they satisfy identically the required
set of fluid equations. The proof of this statement is straightforward and
is given in Appendix B.

\textbf{ACKNOWLEDGEMENTS} 

Work developed in cooperation with the CMFD Team, Consortium for
Magneto-fluid-dynamics (Trieste University, Trieste, Italy).\ Research
partially performed in the framework of the COST Action P17 (EPM, \textit{%
Electromagnetic Processing of Materials}), the GDRE (Groupe de Recherche
Europ\'{e}en) GAMAS and the MIUR (Italian Ministry of University and
Research) PRIN Programme: \textit{Modelli della teoria cinetica matematica
nello studio dei sistemi complessi nelle scienze applicate}.

\section{APPENDIX A - Deterministic description of NS fluids}

In fluid dynamics the state of a fluid is assumed to be prescribed by an
appropriate set of suitably smooth functions $\left\{ Z\right\} \equiv
\left\{ Z_{i},i=1,n\right\} $ denoted as \emph{fluid fields}$.$ These are
required to be real functions of the form
\begin{equation}
Z_{i}=Z_{i}(\mathbf{r,}t),
\end{equation}%
with $\mathbf{r}$ and $t$ spanning respectively the sets $\Omega $
(configuration space) and $I$ (time axis). In particular the functions $%
Z_{i}(\mathbf{r,}t)$ are assumed at least continuous in all points of a
closed set $\overline{\Omega }\times \overline{I}$ (extended configuration
space), with $\overline{\Omega }$ closure of $\Omega ${$.$ In the remainder
we shall require that:}

\begin{enumerate}
\item {$\Omega ${\ is a bounded subset of the Euclidean space }$E^{3}$ on }$%
\mathbb{R}
${$^{3};$}

\item $I$ is identified, when appropriate, either with a bounded time
interval, \textit{i.e.}, $I${$=$}$\left] {t_{0},t_{1}}\right[ \subseteq
\mathbb{R}
,$ or with the real axis $%
\mathbb{R}
$;

\item in the open set $\Omega \times ${$I$} the functions $\left\{ Z\right\}
$ are assumed to be solutions of an appropriate closed set of PDE's, denoted
as \emph{fluid equations;}

\item by assumption, these equations together with appropriate initial and
boundary conditions are required to define a well-posed problem with unique
strong solution defined everywhere in $\Omega \times ${$I$}.
\end{enumerate}

The deterministic description of an incompressible NS fluid is provided by
the fluid fields $\left\{ Z\right\} \mathbf{\equiv }\left\{ \mathbf{V}%
,p\right\} ${, with} $\mathbf{V}(\mathbf{r},t)$ and $p(\mathbf{r},t)\geq 0$
denoting respectively the deterministic fluid velocity and pressure and by {%
the \emph{incompressible NS equations} (INSE)}:
\begin{eqnarray}
\rho  &=&\rho _{o},  \label{1a} \\
\nabla \cdot \mathbf{V} &=&0,  \label{1b} \\
N\mathbf{V} &=&0,  \label{1c} \\
Z(\mathbf{r,}t_{o}) &\mathbf{=}&Z_{o}(\mathbf{r}),  \label{1d} \\
\left. Z(\mathbf{r,}t)\right\vert _{\partial \Omega } &\mathbf{=}&\left.
Z_{w}(\mathbf{r,}t)\right\vert _{\partial \Omega }.  \label{1e}
\end{eqnarray}%
Here $N$ is the NS nonlinear operator
\begin{eqnarray}
&&\left. N\mathbf{V}=\frac{D}{Dt}\mathbf{V}-\mathbf{F}_{H},\right.
\label{2a} \\
&&\left. \frac{D}{Dt}\mathbf{V}\equiv \frac{\partial }{\partial t}\mathbf{V}+%
\mathbf{V}\cdot \nabla \mathbf{V,}\right.   \label{2b} \\
&&\left. \mathbf{F}_{H}\equiv \mathbf{-}\frac{1}{\rho _{o}}\nabla p+\frac{1}{%
\rho _{o}}\mathbf{f}+\upsilon \nabla ^{2}\mathbf{V,}\right.   \label{2c}
\end{eqnarray}%
with $\frac{D}{Dt}\mathbf{V,F}_{H}$ and $\mathbf{f}(\mathbf{r,}t)$ denoting
respectively the Lagrangian fluid acceleration, the total force per unit
mass and the volume force density acting on the fluid, while $\rho
_{o}>0,\nu >0$ the constant mass density and the constant kinematic
viscosity. In particular, $\mathbf{f}$ is assumed of the form%
\begin{equation}
\mathbf{f}=\mathbf{-\nabla }\phi (\mathbf{r},t)+\mathbf{f}_{R},  \label{2d}
\end{equation}%
$\phi (\mathbf{r},t)$ denoting a suitable scalar potential, so that the
first two force terms [in Eq.(\ref{2c})] can be represented as $-\nabla p+%
\mathbf{f}$ $=-\nabla p_{r}+\mathbf{f}_{R},$ with
\begin{equation}
p_{r}(\mathbf{r},t)=p(\mathbf{r},t)-\phi (\mathbf{r},t),  \label{2e}
\end{equation}%
denoting the \emph{reduced fluid pressure}. Moreover, Eqs. (\ref{1a}), (\ref%
{1b}), \ref{1c}), (\ref{1d}) and (\ref{1e}) are respectively the
incompressibility, isochoricity, NS equations and the initial and Dirichlet
boundary conditions for $\left\{ Z\right\} ,$ with $\left\{ Z_{o}(\mathbf{r}%
)\right\} $ and $\left\{ \left. Z_{w}(\mathbf{r,}t)\right\vert _{\partial
\Omega }\right\} $ suitably prescribed deterministic initial and
boundary-value fluid fields, defined respectively at the initial time $%
t=t_{o}$ and on the boundary $\partial \Omega .$

\section{Appendix B - Time evolution of the NS fluid fields}

Let us now show that the phase-space dynamics of thermal tracer particles
determines uniquely the complete set of fluid fields $\left\{ \mathbf{V}%
,p\right\} $. To prove the statement, let us assume that (at the initial
time $t_{o}\in I$) the initial values of the fluid fields $\left\{ \mathbf{%
V(r}_{o},t_{o}),p\mathbf{(r}_{o},t_{o})\right\} $ are prescribed for all $%
\mathbf{r}_{o}\in \overline{\Omega }.$ The let us require that at any time $%
t>t_{o}$ (with $t\in I$) an arbitrary position $\mathbf{r}$ (this includes
also the case in which $\mathbf{r}$ belongs to the boundary $\partial \Omega
$) is reached (at least) by two thermal tracer particles with phase-space
trajectories. Denoting by $\mathbf{x}_{i}(t)=\left\{ \mathbf{r}_{i}(t)=%
\mathbf{r},\mathbf{v}_{i}(t)\right\} $ (for $i=1,2$) their phase-space
states, Eqs.(\ref{EQ.0-a})-(\ref{EQ.2}) then imply that for $i=1,2$ it must
result%
\begin{equation}
\mathbf{v}_{i}(t)=\mathbf{V}(\mathbf{r},t)+\mathbf{n}_{i}(\mathbf{r}%
,t)v_{th}(\mathbf{r},t),  \label{end-1}
\end{equation}%
where the unit vectors $\mathbf{n}_{i}(\mathbf{r},t)$ are, by definition,
solutions of the initial value problem (\ref{EQ.3}) for the initial
conditions $\mathbf{n}(\mathbf{r}_{i}(t_{o}\ ),t_{o})=\mathbf{n}_{i}(\mathbf{%
r}_{oi},t_{o})$ and $v_{th}(\mathbf{r},t)=\sqrt{\frac{2p_{1}(\mathbf{r},t)}{%
\rho _{o}}}.$ If we require, in particular, that $\mathbf{n}_{1}(\mathbf{r}%
,t)=-\mathbf{n}_{2}(\mathbf{r},t)$ the same conclusion is immediate, because
then
\begin{eqnarray}
\mathbf{V}(\mathbf{r},t) &=&\frac{\mathbf{v}_{1}(t)+\mathbf{v}_{2}(t)}{2},
\label{end-2} \\
v_{th}(\mathbf{r},t) &=&\frac{\left\vert \mathbf{v}_{1}(t)-\mathbf{v}%
_{2}(t)\right\vert }{2},  \label{end-3}
\end{eqnarray}%
where the second equation yields $p(\mathbf{r},t)$ in terms of Eq.(\ref%
{kinpressure}). Let us now assume, instead, more generally that there
results $\mathbf{n}_{1}(\mathbf{r},t)\neq -\mathbf{n}_{2}(\mathbf{r},t),%
\mathbf{n}_{2}(\mathbf{r},t).$ In this case it follows%
\begin{eqnarray}
&&\left. v_{th}(\mathbf{r},t)=\frac{\left\vert \mathbf{v}_{1}(t)-\mathbf{v}%
_{2}(t)\right\vert }{\left\vert \mathbf{n}_{1}(\mathbf{r},t)-\mathbf{n}_{2}(%
\mathbf{r},t)\right\vert },\right.  \label{end-4} \\
&&\mathbf{V}(\mathbf{r},t)=\frac{\mathbf{v}_{1}(t)+\mathbf{v}_{2}(t)}{2}-
\label{end-5} \\
&&-\left[ (\mathbf{n}_{1}(\mathbf{r},t)+\mathbf{n}_{2}(\mathbf{r},t)\right]
v_{th}(\mathbf{r},t).  \notag
\end{eqnarray}

Therefore, again, the fluid fields are uniquely determined at any $(\mathbf{r%
},t)\in \overline{\Omega }\times I.$

\end{document}